\documentstyle[jkas]{article}

\beginpage{1}
\endpage{4}

\runningauthor {T. C. Hinse et al.} 
\runningtitle{Orbital stability of SW Lyncis}

\month{May} \year{2014} \volume{XX} \issueno{YY}
\beginpage{1}\endpage{9}

\date{Received ... 2014; Accepted ... 2014}

\begin{document}

\title{An Orbital Stability Study of the Proposed Companions of SW Lyncis} 

\author{T. C. Hinse$^{1}$, Jonathan Horner $^{2,3}$, Robert A. Wittenmyer$^{3,4}$} 
\address{$^1$ Korea Astronomy and Space Science Institute,
776 Daedukdae-ro, Yuseong-gu, Daejeon 305-348, Republic of Korea.\\
 {\it E-mail : tchinse@gmail.com}}

\address{$^2$ Computational Engineering and Science Research Centre, University of Southern Queensland, Toowoomba, Queensland 4350, Australia.}

\address{$^3$ Australian Centre for Astrobiology, University of New South 
Wales, Sydney 2052, Australia.}

\address{$^4$ School of Physics, University of New South Wales, Sydney 2052,
              Australia.}

\address{\normalsize{\it (Received ... 2014; Accepted ... 2014)}}
\offprints{T. C. Hinse}

\abstract{We have investigated the dynamical stability of the proposed companions orbiting the Algol type short-period eclipsing binary SW Lyncis (Kim et al. 2010). The two candidate companions are of stellar to sub-stellar nature, and were inferred from timing measurements of the system's primary and secondary eclipses. We applied well-tested numerical techniques to accurately integrate the orbits of the two companions and to test for chaotic dynamical behaviour. We carried out the stability analysis within a systematic parameter survey varying both the geometries and orientation of the orbits of the companions, as well as their masses. In all our numerical integrations we found that the proposed SW Lyn multi-body system is highly unstable on time-scales on the order of 1000 years. Our results cast doubt on the interpretation that the timing  variations are caused by two companions. This work demonstrates that a straightforward dynamical analysis can help to test whether a best-fit companion-based model is a physically viable explanation for measured eclipse timing variations. We conclude that dynamical considerations reveal that the propsed SW Lyncis multi-body system most likely does not exist or the companions have significantly different orbital properties as conjectured in Kim et al. (2010).}


\maketitle

\section{INTRODUCTION}
In recent years, a number of multiple star systems have been proposed to orbit a binary pair as a result of photometric follow-up observations of eclipsing binaries. The nature of the proposed companions ranges from planetary to sub-stellar objects. \citet{Lee2009} were the first to propose such circumbinary companions to explain eclipse timing variations, suggesting that the short-period pulsating subdwarf binary HW Virginis (hereafter HW Vir) was being accompanied by two unseen circumbinary companions of mass 8.5 $M_{jup}$ and 
19.2 $M_{jup}$. Following this announcement, \citet{Beuermann2010} announced 
the detection of two planets with masses of 2 $M_{jup}$ and 7 $M_{jup}$ 
orbiting the recently formed post-common envelope binary NN Serpentis (hereafter NN Ser). \citet{Qian2011} then also announced the discovery of two circumbinary companions orbiting the eclipsing polar HU Aquarii (herafter HU Aqr) using the same detection technique. Furthermore, \cite{Potter2011} announced a possible detection of two giant extrasolar planets orbiting the eclipsing polar UZ Fornacis (hereafter UZ For) and \cite{Lee2012} found periodic signatures from photometric measurements of the Algol system SZ Herculis (hereafter SZ Her). Here, the authors associated their detection with the possible existence of two M-type stellar companions with minimum massess 0.19 $M_{\odot}$ and 0.22 $M_{\odot}$. Finally, a recent study by \cite{Almeida2013} also proposed the existence of two circumbinary companions orbiting the post-common envelope binary NSVS 14256825. In addition to these proposed systems, a number of studies (e.g. AH Cephei in \citet{Kim2005}) proposed unseen companions around several other close binaries in the last few years and we refer to \citet[and references therein]{Hinse2014a} for further details of these systems.

The observational technique which is used for the discovery of possible circumbinary companions is primarily based on the timing measurements of the primary eclipse of the binary star system. Considering the case when the primary is isolated and positioned at a constant distance to Earth, the time of primary eclipses in the future will follow a linear ephemeris relative from some reference epoch $T_0$ with binary period $P_0$. However, if an additional massive companion is gravitationally bound to the binary components, then the binary system will start to follow an orbital trajectory around the total system barycenter. This gives rise to the so-called light-travel time effect (LTTE) \citep{Irwin1952, Borkovits1996}. As a consequence of the finite speed of light, the arrival-time of photons will be delayed (advanced) as a result of the distance between the binary and the Earth being a maximum (minimum). The manifestation of this effect is a quasi-periodic change in the measured timings of the primary eclipses and is also known as eclipse timing variations (ETV). The precision with which timing measurements are obtained is mainly governed by the photometric quality of the data, the observing cadence during the eclipse, and the presence of star-spots. In general, timing measurements should be independent of spectral band observations although datasets with mixed timing measurements obtained from various filters could result in systematics uncertainties and possibly lead to false interpretation of the period variations of an eclipsing binary \citep{Gozdziewski2012}.

For nearly all systems with a proposed circumbinary companion, as mentioned above, there is a fundamental problem that raises doubts towards the correct interpretation of the measured eclipse timing variations. A common denominator for all systems is the three-body problem: two massive companions orbiting a binary star. From a dynamical point of view, such configurations naturally 
raise the question of orbital stability. The numerical demonstration of a long-lived stable three-body system could serve to further support the interpretation of observed timing variations as being directly caused by the perturbing effects of massive companions. One other possibility is that the period variations are indeed caused by additional companions, but in this case the orbital architecture must be significantly different than discussed.

As an example the only multi-body system that seem to follow stable orbits around a post-common envelope (evolved) binary is the NN Ser system \citep{Beuermann2010,Horner2012a,Beuermann2013}. Recently the planetary interpretation of the primary eclipse times of NN Ser was further supported by timing measurements of the secondary eclipses. \citet{Parsons2014} were able to rule out the possibility of apsidal motion of the orbit of NN Ser showing that the secondary eclipse timings followed the same trend as the primary timing measurements. Furthermore, a stable multi-body circumbinary system was recently detected using \texttt{Kepler} space-telescope data. \citet{Orosz2012} utilized the transit detection technique to detect two planets transiting a main-sequence primary star very similar to the Sun accompanied be a cooler M-type secondary.

In contrast to the stability and feasibility demonstrated for the two systems discussed in the previous paragraph, dynamical studies of the other circumbinary systems discussed above have instead revealed a very different picture. Rather than featuring proposed companions that move on dynamically stable orbits, the companions in those systems have instead been found to move on highly unstable orbits (with the exception of NN Ser). Typically, the companions will either experience close encounters resulting in the ejection of one or both components or there will be direct collision events. Several studies have recently focussed on the orbital stability of the proposed circumbinary systems. The first study to test for the orbital longevity of any such post-common eclipsing binary system (HU Aqr) was presented in \citet{Horner2011}. In their work they followed the orbits as part of a detailed dynamical analysis and demonstrated that the proposed two-planet system would be highly unstable, with break-up time-scales of less than a few thousand years. Two follow-up studies of HU Aqr were recently presented \citep{Hinse2012a, Wittenmyer2012}. In the first, where the authors attempted to determine new best-fit models to the observed timing data accompanied with orbital stability requirements. 

In that work, the authors found a new orbital architecture for the proposed companions around HU Aqr, but again found that architecture to be highly unstable. Long-lived orbits capable of surviving on million-year timescales were only found for HU Aqr when additional orbital stability constraints were imposed on an ensemble of best-fit solutions, based on the Hill radii of the proposed companions. The key difference between the stable solutions found in this manner and the unstable ones that resulted solely from the observational data was that the stable solutions featured near-circular orbits for the two companions. Two further studies of the HU Aqr system \citep{Gozdziewski2012,Wittenmyer2012} both suggested that two-companion solutions could be ruled out for the system, with \citet{Gozdziewski2012} pointing towards an alternative, single companion model as providing the best explanation of the measured timing variations."

\begin{table*}[t]
\begin{center}
\centering
\begin{tabular}{cccc} 
\hline \hline
Element & Linear term & Inner LITE orbit & Outer LITE orbit \\
\hline
$T_0$ (HJD) & 2,443,975.3869(1) & - & - \\
$P_0$ (days) & 0.64406637(2) & - & - \\
$a\sin I$ (au) & - & 1.333(9) & 0.742(18) \\
$e$ & - & 0.581(6) & 0.00(3) \\
$\omega$ (degrees) & - & 188(7) & - \\
$T$ (HJD) & - & 2,438,818(12) & - \\
$P$ (years) & - & 5.791(4) & 33.9(5) \\
\hline
$K$ (days) & - & 0.0063(1) & 0.0043(3) \\
$m\sin I (M_{\odot})$ & - & 0.91(2) & 0.14(1)\\
\hline 
\end{tabular}
\caption{Best-fit elements (the first 7) of the two LITE orbits (determined from simultaneous fitting) as reproduced from \cite[their table 3]{Kim2010}. 
$K$ measures the semi-amplitude and is calculated from Eq. 4 in \cite{Irwin1952}. The minimum masses for the two companions is determined iteratively using the mass-function and are consistent with two separate Kepler orbits with the combined binary in one focus of the ellipse. Because the 4th body orbit is circular $(e_2=0.0)$ the argument of pericenter $(\omega)$ and time of pericenter passage $(T)$ are undefined. Numbers in paranthesis denote the uncertainty of the last digit as adopted from \cite{Kim2010}. The mass of the primary and secondary components are $1.77~M_{\odot}$ and $0.92~M_{\odot}$, respectively.}
\label{Table1}
\end{center}
\end{table*}

Several additional studies exist that demonstrate orbital instability and/or unconstrained orbital parameters of proposed multi-body circumbinary systems (HW Vir, SZ Her, QS Vir, NSVS 14256825, RZ Dra) and we refer the reader to the following sources in the literature \citep{Horner2012b, Hinse2012b, Horner2013, Wittenmyer2013, Hinse2014a, Hinse2014b} for more details.

In this work we present the results of a dynamical orbit stability analysis of the two proposed circumbinary companions of the eclipsing binary SW Lyncis (hereafter SW Lyn, \citet{Kim2010}). In their timing analysis of historical plus newly acquired photometric observations the authors find sound evidence of a 5.8-year cycle along with a somewhat less tighly constrained cycle of 33.9 years. In their quest to find a plausible explanation the authors attempted to explain (among other possible explanations) the timing variations with a possible pair of light-travel time orbits corresponding to two circumbinary companions. In their discussion on the 34-year cycle 
\cite{Kim2010} highlight that the two conjectured companions are unlikely to approach each other within 5 au when considering co-planar orbits. This statement motivated us to test the system's overall stability by numerically evaluating the orbital trajectories using the osculating best-fit Keplerian parameters (as obtained from their light-travel time model) and their corresponding errors \citep{Kim2010} as the initial conditions in this work.

This work is structured as follows. In section 2 we briefly review the mathematical formulation of the LITE effect resulting in the proposition of the two possible circumbinary companions. We highlight the underlying assumptions and also outline the derivation of the companion orbits from their associated LITE orbits. In section 3 we give a short description of the numerical techniques and methods used in this work. In section 4 we present numerical results of an orbital stability analysis for co-planar companion orbits. In section 5 we generalise and consider scenarios where the two companions move 
on mutually inclined orbits, as well as scenarios in which their orbits are co-planar, but their masses differ from those used in section 4. Section 6 concludes our analysis.

\section{Details of LITE and Orbital Properties of SW Lyn and Proposed Companions}

The mathematical formulation of the single-companion LITE effect was described in great detail by \citet{Irwin1952}. In its simplest version, the modelling of timing measurements requires a set of 2 + 5 parameters. The first two parameters describe the linear ephemeris of the eclipsing binary and the remaining five describe the size, shape and orientation of the LITE orbit. We remind the reader that the LITE orbit utilizes the two-body formulation and represents the orbit of the binary barycenter around the binary-companion center of mass. The first assumption in this formulation is that the binary orbit is small compared to the LITE orbit. A period variation due to LITE is then regarded as a geometric effect. In that case the binary is treated as a single massive object with mass equal to the sum of masses of the two components.

If $T_0$ is chosen to be some arbitrary reference epoch, $P_0$ measures the eclipse period of the binary, and considering the case of a single companion, then the times of primary eclipses at epoch $E$ are given by

\begin{equation}
T(E) = T_{0} + P_{0}\times E + \tau,
\end{equation}
\noindent

where $\tau$ measures the LITE effect and is a function of the orbital elements denoted as projected semi-major axis $(a\sin I)$, eccentricity $(e)$, argument of pericenter $(\omega)$, time of pericenter passage $(T)$ and orbital period $(P)$. The cycle number $E$ appears implicitly via Kepler's equation and we refer the interested reader to \citet{Hinse2012a} for details. In practice, once a best-fit single LITE oribit has been determined, the quantity $T(E) - (T_{0}+P_{0}*E)$ is plotted 
and most often denoted as $"O - C"$, eventually revealing one or more modulations of the binary period.

In the case of a second cyclic variation one often assumes the principle of superposition. Assuming the absence of mutual gravitational interactions between the two companions, the standard praxis in timing analysis-work usually considers two separated Keplerian orbits. Only the interaction between the companion and the combined binary mass is considered. Perturbations between the two unseen companions are neglected. The basis of a timing analysis then attempts to explain the total timing variation as the sum of two LITE orbits 
\citep{Hinse2012b}. All measurements are then simultaneously modelled during the least-squares minimisation procedure with possible weights. In Table \ref{Table1} we reproduce the Keplerian SW Lyn LITE elements from \cite{Kim2010} for the two companions along with their 1-sigma formal uncertainties. We would like to highlight that these orbital elements were calculated neglecting any possible influence of external gravitational perturbations.

\subsection{The SW Lyncis System}

We will now direct our attention to the details of the binary and its proposed companions. SW Lyn is a detached eclipsing binary of Algol type with an orbital period of around 16 hours. The mass of the two components are 1.77 $M_{\odot}$ and 0.92 $M_{\odot}$ \citep{Kim2010}. From Table \ref{Table1} we note that the short-period LITE orbit has an eccentricity of 0.58. The long-period LITE orbit is circular. The masses of the two components are found from the mass-function \citep{Hinse2012a,Hinse2012b} and are therefore minimum masses with the 
$\sin I$ factor undetermined. The geometric orientation of the system can only be definitively determined in the case where the unseen companion is observed to eclipse or transit one or other of the binary components. In that case, it becomes possible to determine the true mass of the unseen companion. In all other cases, the degeneracy between the mass and inclination of the system remains.

The Keplerian orbital elements of a companion can be derived from first principles. As a result of barycentric orbits and as pointed out in 
\cite{Hinse2012b} the eccentricity, the time of pericenter passage and the orbital period of the LITE orbit are the same for the associated companion orbit. Since the two orbits are in the same plane (not to be confused with the two companion orbits) the $\sin I$ factor are also the same. The only differences occur for the argument of pericenter and projected (or minimum) semi-major axes. With respect to the argument of pericenter the  orbit of the companion is anti-aligned to its associated LITE orbit. We therefore have a 
180 degrees difference between the two apsidal lines. The semi-major axis can be computed using one of two different methods. The first method makes use of Kepler's third law. Since the minimum mass of the companion and its orbital period are known quantities the projected semi-major axis of the companion's orbit is given as

\begin{table*}[t]
\begin{center}
\centering
\begin{tabular}{cccc} 
\hline \hline
Element & SW Lyn(AB)C ($i=1$) & SW Lyn(AB)D ($i=2$) \\
\hline
$a_{1,2}\sin I$ (au) from Eq.~\ref{eq1} & $4.954 \pm 0.012$ & $14.816\pm 0.17$\\
$a_{1,2}\sin I$ (au) from Eq.~\ref{eq2} & $5.153 \pm 0.062$ & $14.361\pm 1.09$\\
$e_{1,2}$ & $0.581 \pm 0.006$ & $0.00 \pm 0.030$ \\
$\omega_{1,2}$ (degrees) & $188 \pm 7 - 180 = 8 \pm 7 $ & - \\
$P_{1,2}$ (days) & $2115.16 \pm 1.46$ & $12381.98 \pm 182.6$ \\
$m_{1,2}\sin I (M_{\odot})$ & $0.91 \pm 0.02$ & $0.14 \pm 0.01$ \\
\hline 
\end{tabular}
\caption{Astrocentric orbital elements of the two proposed companions derived from first principles and Kepler's third law of orbital motion. The dynamical center corresponds to the binary barycenter with mass 2.69 $M_{\odot}$. Uncertainties for the derived quantities have been obtained from standard error propagation assuming uncorrelated uncertainties.}
\label{Table2}
\end{center}
\end{table*}

\begin{equation}
a \sin I = \Big( \frac{P^2 (M+m\sin I) k^2}{4\pi^2}\Big)^{1/3},
\label{eq1}
\end{equation}
\noindent
where $M$ is the total mass of the dynamical center which in this case is the combined binary components ($M = 2.69~M_{\odot}$) and $k^2$ is the Gauss gravitational constant. We would like to stress that the projected semi-major axis $a \sin I$ is relative to the combined binary treated as the dynamical center. Therefore the semi-major axis in the above equation is for an astrocentric system since Kepler's third law is only valid in a system with a single dynamical center.

\begin{figure}[!t]
\centering \epsfxsize=8cm 
\epsfbox{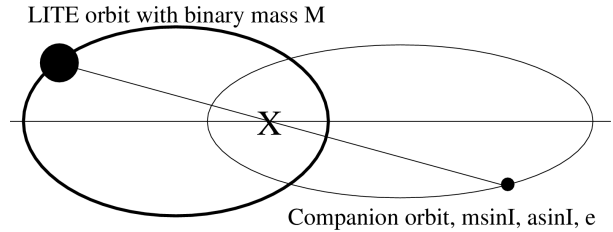} 
\caption{Illustration of the two-body problem in a barycentric reference system. The barycenter is marked with a "X". The (combined) binary has mass M and its orbit corresponds to the LITE orbit. The unseen companion has mass m.}
\label{Figure1}
\end{figure}

The second method considers the two orbits in their barycentric reference frames. To illustrate the difference between these two techniques, in Fig.~\ref{Figure1} we plot a LITE orbit and its associated companion orbit as an example. Following \cite{MurrayDermott2001} the projected semi-major axis of the LITE orbit $(a_{1,LITE}\sin I)$ and the astrocentric orbit $(a_1\sin I)$ of the unseen companion are related to each 
other via the masses as follows

\begin{equation}
a_{1}\sin I = a_{1,LITE}\sin I \frac{m_1 \sin I + M}{m_1 \sin I}.
\label{eq2}
\end{equation}
\noindent

The right-hand side only contains known quantities listed in 
Table~\ref{Table1}. In Table~\ref{Table2} we show numerical values of all known orbital quantities for the orbit of the two companions. The semi-major axis as computed from the two methods agree well with the discrepancies (at the 1\% level) most likely resulting from the uncertainties of the best-fit parameters. At this stage we point out that throughout this study we adapt numerical values for the astrocentric orbits (Table \ref{Table2}) as calculated by Eq. \ref{eq1}.

\begin{figure}[!t]
\centering \epsfxsize=8cm 
\epsfbox{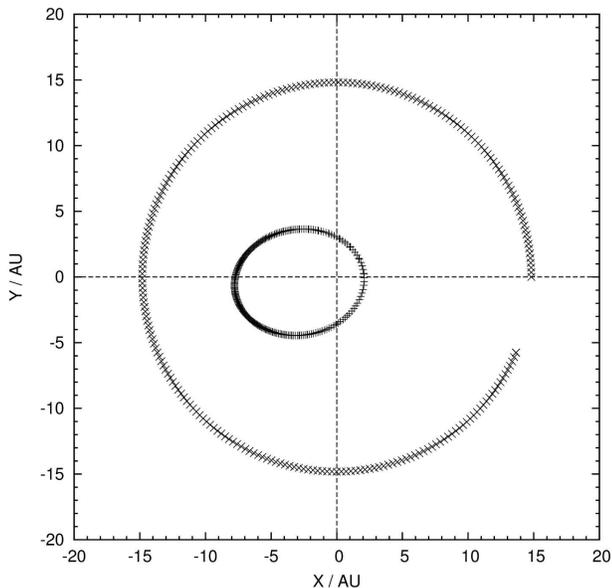} 
\caption{Geometry of the two unseen companions. Here we have projected their orbits on the skyplane with North being up and East being left. Both orbits were integrated numerically within the framework of the two-body problem. The origin of the coordinate system is the (approximate) barycenter of the binary and companion. The outer orbit is plotted for almost one orbital period.}
\label{Figure2}
\end{figure}

In Fig.~\ref{Figure2} we show two Keplerian orbits of the companions perpendicular to the sky plane assuming co-planar orbits $(I = I_1 = I_2 = 90^{\circ})$. The orbital apocenter of the inner orbit is calculated as 
$a_1(1 + e_1) = 4.9(1+0.581) = 7.7$ au. This distance implies that the two orbits are well separated and is a promising indication of stability. However, the masses of the proposed SW Lyn companions are large, and might render the system too energetic for their orbits to remain gravitationally bound on long timescales. This should be tested. The interesting question is whether the orbits will remain relatively unperturbed and continue to trace out their respective paths in a numerical integration? To support and substantiate the LITE interpretation of the timing measurements as a periodic recurring phenomena due to two massive companions, the answer should be yes. A dynamical analysis will be the subject of the next sections considering various orbital geometries as well as different masses of the companions to infer the dynamical stability of the SW Lyn multi-body system.

\section{Orbit Integration Technique and Numerical Methods}

A dynamical analysis aims to investigate the temporal evolution of an ensemble of orbits located in the neighbourhood of the best-fit solution. In this work we utilise two distinct numerical methods. The first technique involves the orbit integration package \texttt{MERCURY} \citep{Chambers1999}. This package allows the numerical integration of single orbits gravitationally interacting with each other. It offers several algorithms for the solution of the first order differential equations describing the system's equations of motion. In this work we made use of the Bulirsch-Stoer method featuring adaptive time stepping to accurately resolve close encounters. In all our integrations we used an initial time step of 0.01 days. The integration accuracy parameter was set to $10^{-14}$. The package allows the specification of initial conditions in an astrocentric reference frame and is therefore suitable for our problem. We have previously applied this package in similar studies and we refer to \citet{Horner2011, 
Hinse2012b} and \citet{Hinse2014a} for numerical tests.

The other technique is the computation of a fast chaos indicator known as MEGNO (Mean Exponential Growth factor of Nearby Orbits) as introduced by \cite{Cincotta2003}. The latter found wide-spread application in dynamical astronomy and celestial mechanics \citep{Gozdziewski2001,Hinse2010,Kostov2013} and is an effective tool to explore the phase-space topology of a dynamical system. In this work we have applied the MEGNO technique to the gravitational three-body problem with focus on the proposed companions around SW Lyn.  Our computations have made use of the KMTNet\footnote{Korea Microlensing Telescope Network} computing cluster (multi-core super-computer using 33 Intel Xeon X5650 cores each running at 2.7 GHz) to compute the dynamical MEGNO maps using the newly developed MECHANIC \citep{Slonina2015} single task-farm software package.

The details of MEGNO are as follows. For a given initial condition of the three-body problem the equations of motion and variational equations \citep{MikkolaInnanen1999} are solved in parallel. The MEGNO, usually denoted 
as $\langle Y \rangle$, is then computed as described in detail in  \citet{Gozdziewski2001}. In brevity, if $\langle Y \rangle$ after some integration time remains close to $\langle Y \rangle = 2$, then the orbit is characterised by a quasi-periodic time evolution. However, if $\langle Y \rangle$ is significantly larger than 2, we then judge the orbit to be chaotic. For clarity, a chaotic system does not automatically imply unstable orbits. However, unstable orbits will always imply chaotic time evolution. The important key-issue to consider is the integration length. If the moment of chaotic onset in the dynamical system requires a much longer time period than the integration time, then the possibility of erroneously concluding quasi-period is real. Therefore, one should integrate the system for long enough in order to allow the system to possibly exhibit chaotic behaviour. In this case, we find that integrating the SW Lyn three-body system for 5000 orbits of the inner companion spans a sufficiently long time period to allow us to make firm conclusions on the overall stability of the SW Lyn multi-body system.

\section{Orbital Stability Analysis - Coplanar Orbits}

A fundamental unknown is the orbital orientation $(\sin I)$ of the LITE orbit allowing us to only determine the minimum mass of the unseen companions. 

We first considered the most simple solution for the orbital geometry of the two unseen companions - co-planar orbits (following our earlier work; e.g. \citet{Horner2011,Hinse2012a,Hinse2012b,Wittenmyer2012,Wittenmyer2013}). The assumption of co-planar orbits is reasonable given that any companions would most likely have formed from a single circumbinary protoplanetary disk. In all calculations the binary was treated as a single massive object in order to be consistent with the LITE formulation. Initial conditions for the two companions are listed in Table~\ref{Table2}. The uncertainties in projected semi-major axis were obtained from standard error propagation.

\begin{figure}[!t]
\centering \epsfxsize=8cm 
\epsfbox{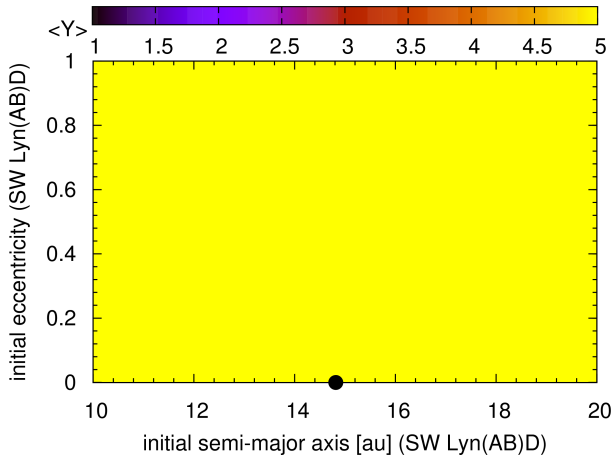} 
\caption{Dynamical MEGNO map for the outer companion of SW Lyn. Because the orbital parameters of the inner companion are well determined we kept them fixed at their osculating values shown in Table~\ref{Table2}. The black dot indicates the best-fit osculating orbit of SW Lyn(AB)D.}
\label{Figure3}
\end{figure}

\begin{figure*}[!t]
\centering \epsfxsize=17cm 
\epsfbox{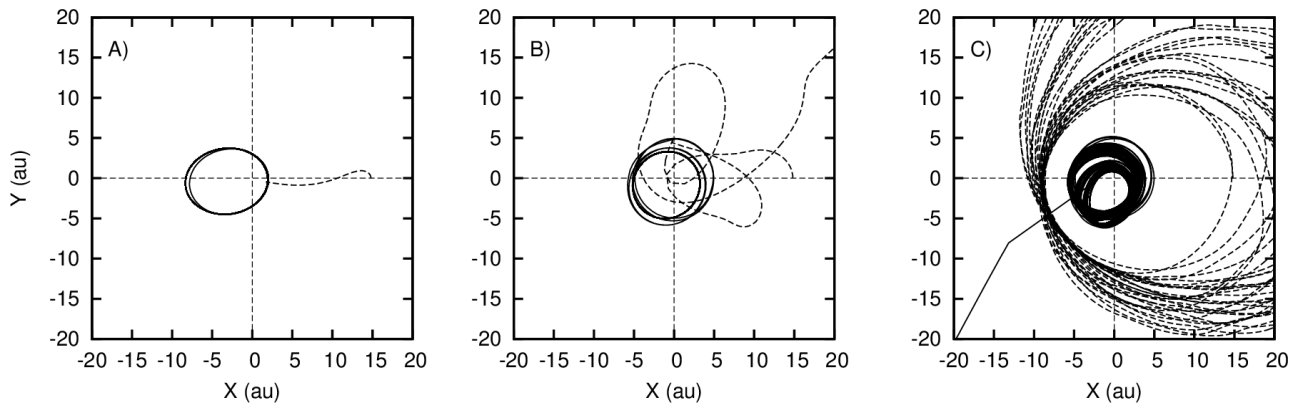} 
\caption{Results from direct integrations of the SW Lyn three-body problem for $\sin I = 1$. We consider three cases. Panel A): Initial conditions as shown in Table~\ref{Table2}. Panel B): Same as previous panel, but now the eccentricity of inner companion is set to zero (circular orbit). Panel C): Same as previous panel, but now the mass of inner companion is set to 0.14 $M_{\odot}$.}
\label{Figure4}
\end{figure*}

\begin{figure*}[!t]
\centering \epsfxsize=17cm 
\epsfbox{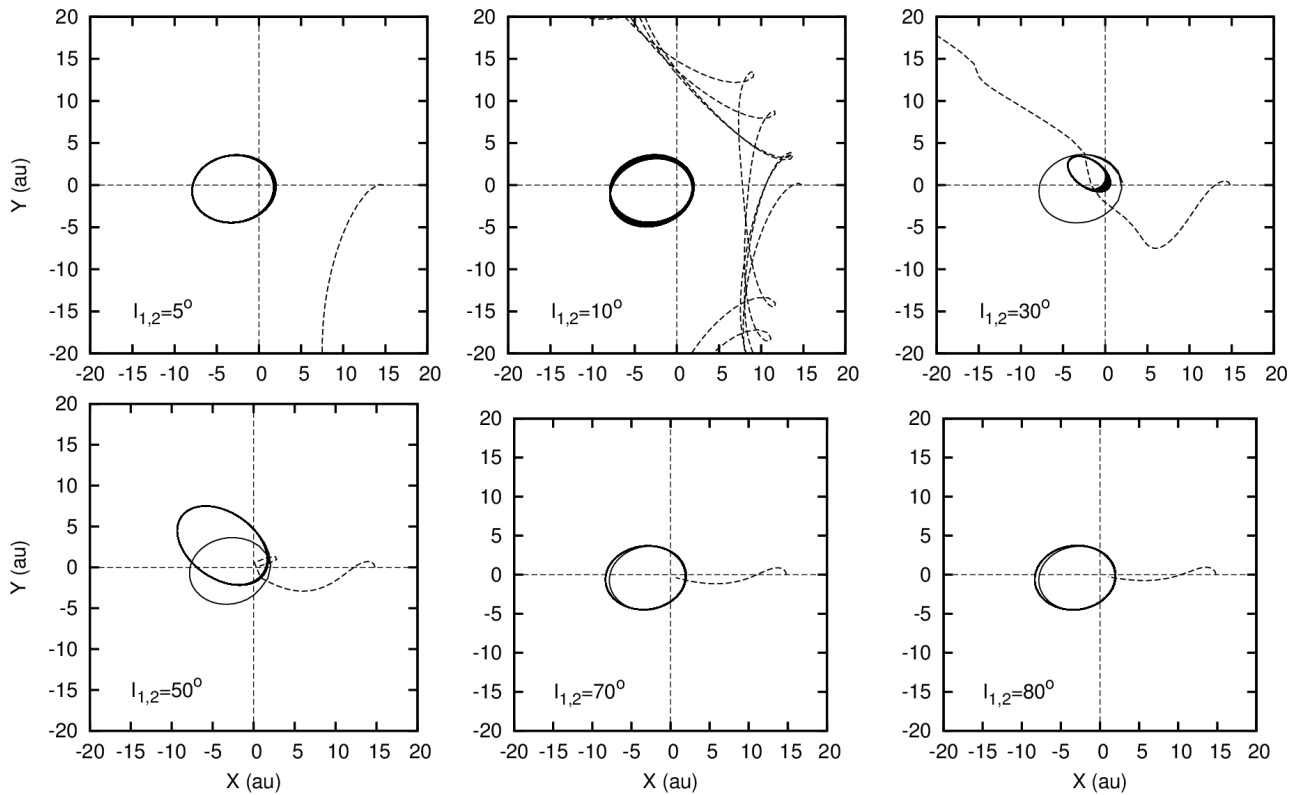} 
\caption{Results from direct integrations of the SW Lyn three-body problem considering scenarios in which the orbits of the unseen companions are coplanar, but aligned at varying angles to our line of sight. The mass of the two companions were scaled accordingly. The two companions are still embedded in the same plane. Both ejections and collisions events were registered shortly after the start of integration. All initial conditions follow highly unstable orbits. The masses for the two companions were as follows. 
$I_{1,2} = 5$: (inner=$10.44~M_{\odot}$, outer=$1.61~M_{\odot}$). 
$I_{1,2} = 10$: (inner=$5.24~M_{\odot}$, outer=$0.81~M_{\odot}$). 
$I_{1,2} = 30$: (inner=$1.82~M_{\odot}$, outer=$0.28~M_{\odot}$). 
$I_{1,2} = 50$: (inner=$1.19~M_{\odot}$, $0.18~M_{\odot}$). 
$I_{1,2} = 70$ (inner=$0.97~M_{\odot}$, $0.15~M_{\odot}$). 
$I_{1,2} = 80$: (inner=$0.92~M_{\odot}$, outer=$0.14~M_{\odot}$).}
\label{Figure5}
\end{figure*}

We first calculated a dynamical MEGNO map exploring the $(a_2,e_2)$ space of the outer companion. We considered a large range in orbital semi-major axis and eccentricties. Since the orbit of the short-period companion is relatively well characterised, we kept its orbit fixed. The result is shown in Fig.~\ref{Figure3}. We explored the range $a_{2} \in [10, 20]$ au and $e_2 \in [0,1]$. For all probed orbits we find the system to exhibit a chaotic time evolution. 

However, in the astrodynamical multi-body problem a chaotic orbit does not stricly imply instability. We therefore investigated the stability of single orbits by considering a large ensemble of initial conditions within the 1-sigma error uncertainties of the orbital parameters. In each integration the system was followed for 10000 years. We investigated the effects of placing the proposed companions at different initial mean longitudes considering the range [0,360] in steps of 10 degrees. For the inner eccentric orbit we also investigated the influence of the argument of pericenter parameter by also considering the range [0,360] of this angle in steps of 10 degrees. This was not possible for the outer companion as the best-fit LITE orbit seems to be very circular. Hence the argument of pericenter is not defined. Systematic combinations of those angles were also considered and tested as part of our stability study. In addition we also varied the mass and eccentricities of the two companions. 

\begin{figure*}[!t]
\centering \epsfxsize=17cm 
\epsfbox{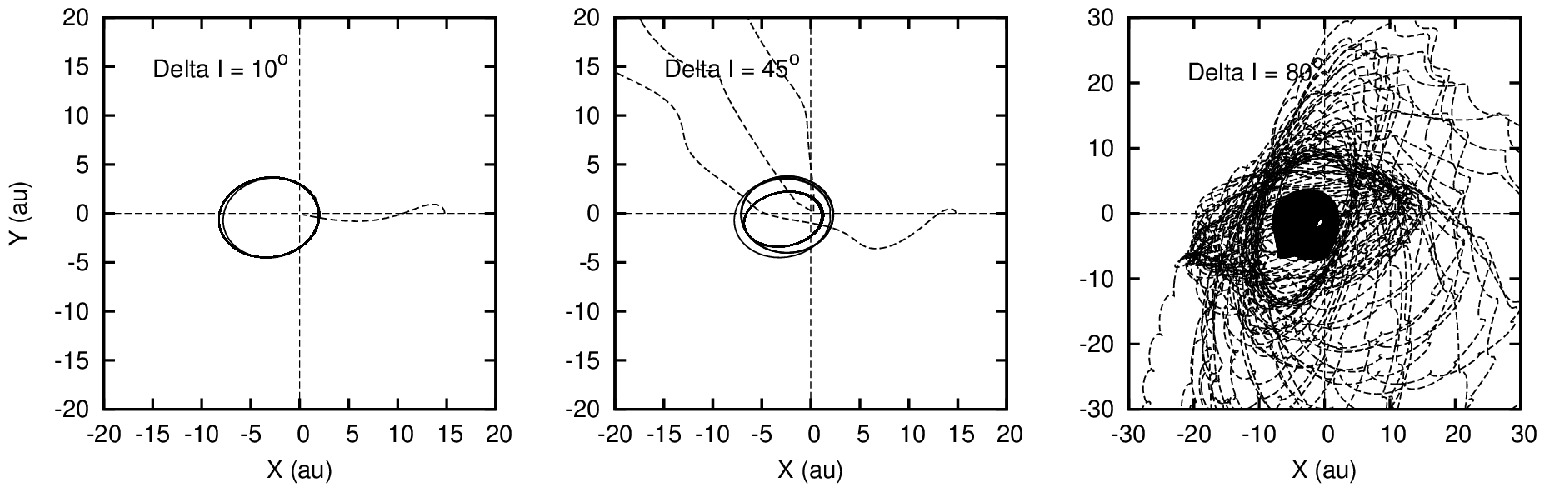} 
\caption{Results from direct integrations of the SW Lyn three-body problem considering mutually inclined orbits. The three panels show the orbits with relative inclinations of 10, 45 and 80 degrees. The masses of the companions were taken to be their minimum values as shown in Table 2.}
\label{Figure6}
\end{figure*}

In all cases we found the system to be highly unstable with one of the components either being ejected from the system or colliding with the central binary. To illustrate our findings we show some results in Fig.~\ref{Figure4}. For Fig.~\ref{Figure4}A and Fig.~\ref{Figure4}B the inner companion collided with the central binary after just a few years. For Fig.~\ref{Figure4}C the orbit survived for 1000 years. However, their time evolution obviously does not resemble the geometry of the two proposed companions as presented in \cite{Kim2010}. In fact, this system is unstable in the sense that the outer companion collided with the central binary after 3704 years and the inner companion was ejected after 3281 years. We show these particular examples as the considered parameters should render the system to become more stable. In general low-mass and circular orbits will always have the effect to increase the longevity of a gravitational multi-body system. The solutions highlighted in this figure were chosen as they represent cases where the initial orbital parameters should have been the most promising in terms of the stability of the system - with low eccentricities and masses for the companion bodies.

\section{Orbital Stability Analysis - Inclined Orbits}

We have also considered various inclinations of the orbits relative to the sky-plane. The orbits were still considered to be coplanar relative to each other. We have therefore considered several values of the line-of-sight to sky-plane inclinations and scaled the masses accordingly for the two companions. However, we stress that the most likely geometric orientation are orbits with $\sin I = 90^{\circ}$ since the companions were most likely formed in the same plane as the binary orbit. An example of such a system would be Kepler-16 \citep{Doyle2011} consisting of a transiting circumbinary planet embedded in the same plane as the binary orbit. In Fig.~\ref{Figure5} we show the results from our survey. Initial conditions for all the other orbital parameters are shown in Table~\ref{Table2}. The results rigorously show that all considered start conditions result in unstable orbits. As was the case in the previous section we detected both collision and ejection events. In particular, ejection events are clearly demonstrated for $I_{1,2} = 5, 10, 30, 50$ degrees. The remaining two cases ($I_{1,2} = 70$ and $I_{1,2} = 80$ degrees) resulted in a collision between the outer companion and the central binary.

A final exercise in this stability study was to consider mutual inclinations between the two companions. Invoking a relative inclinations reflects the situation where the two companions have not formed from the same disk or their orbits have subsequently evolved as a result of unknown perturbations (i.e Kozai cycles due to a distant massive perturber) leading to a non-coplanar configuration. We have considered several relative inclinations and retained the mass of the two companions to be their minimum mass values. We considered several values of mutual orbital inclination, and in each case gave the system the maximial likelihood of stability by setting the mass of both companions to their minimum mass values. A subset of our test orbits are plotted in Fig.~\ref{Figure6} considering three values of the relative inclination. Again, we find that the orbits tend to be highly unstable, and diverge from those proposed for the two companions on the basis of LITE analysis on very short timescales, drawing significant doubt on the currently proposed nature of the system.

\section{Conclusions and Discussion}

In this study we have carried out a detailed orbital stability study of the multi-body system proposed to orbit around SW Lyn. In their work 
\cite{Kim2010} conjecture about the possibility of the existence of two circumbinary companions forming a quadruple system. The authors present substantial modelling work that aims to explain the observed timing variations by a pair of light-time orbits while pointing out that the outer companion might be doubtful. In this work we have rigorously showed that all our numerical integrations resulted in a swift disintegration of the proposed system, with the unseen companions being removed through collision or ejection on timescales of just a few thousand years, or less. This allows us to conclude that the  proposed companions most likely do not exist or the companions exhibit a much different orbital architecture.

Several assumptions were made and in the following we would like to discuss some of them. First the mathematical formulation of the LITE effect assumes that the binary can be replaced by a single massive object positioned at the binary barycenter. This assumption might be acceptable provided that the companion orbits are much larger than the binary orbit. Otherwise, gravitational perturbations on the binary orbit will result in additional eclipse timing variation. {\bf Furthermore, all objects in this study were treated as pointmasses. This implies that we have not considered tidal effects between otherwise extended masses. However, at current time we are not aware of the possibility that tidal interaction could have a significant stabilising effect on the orbits of gravitationally interacting bodies. This possibility is an interesting question and might form part of a future investigation. A detailed treatment of tidal interaction is beyond the scope of this study.} 

Another assumption is the application of the superposition principle of two light-time orbits. In principle, this approach is incorrect, since the two companions will clearly perturb one another's orbits. This in turn, would introduce a feedback to the binary orbit, which will also change as a result, driving more complex timing variation in addition to the geometric LITE effect. The effects of mutual interactions are important to take into account, especially when considering sub-stellar mass companions on slight to moderate eccentric orbits.

However, for smaller masses the principle of superposition applied to two light-time orbits is more correct as the two masses interact less with each other. This situation has recently been demonstrated through the generation of synthetic $n$-body data aiming to model the light-travel time effect caused by two interacting circumbinary planets \citep{Hinse2014c}. These authors numerically created a synthetic dataset which mimics a two-body light-travel time effect. They successfully reproduced the known input parameters of the two planets from a least-squares minimisation technique.

Recently, the LITE effect has been formulated in terms of Jacobi coordinates and might serve as an alternative to the superposition principle 
\citep{Gozdziewski2012}. In their work the authors describe the LITE orbit as a result of several companions in a hierarchical order.

In this work, we have explicitly shown observed eclipse timing variations of the SW Lyn system can not be the result of the unseen massive companions proposed by Kim et al. (2010). The system conjectured in that work proved to be unstable on timescales of just a few thousand years - far too short to be considered dynamically feasible. One possible explanation for this discrepancy might be that the observed 34-year modulation of the system may not be the result of an additional companion, but may instead have another explanation. In other words, the short-period modulation may well turn out to be the result of an unseen companion, with the long-period trend instead being the result of the magnetic activity of one or other of the binary components. Such a one-companion interpretation could well be more viable, since it essentially solves the instablity issue. Indeed, such a solution was recently proposed to explain the observed timing variations of the HU Aqr binary system \cite{Gozdziewski2012}. In that work, the authors suggested that mixed timing measurements obtained from different photometric passbands (different spectral domains) might result in unaccounted correlated (red) noise in the timing data. 
The detection of a single LITE orbit with significant confidence was recently announced \citep{Lee2013}. However, if the short-period modulation is truly associated with a companion in SW Lyn, then why does it have a large eccentricity? Large eccentricities are usually explained by the gravitational influence by additional bodies and would point towards an outer companion. Therefore, an additional explanation would be that the period modulation is due to two companions, but exhibiting a substantial different orbital architecture than found by Kim et al. (2010).

The questions concerned SW Lyn are far from answered and future observations will contribute towards a better understanding of this interesting system. Additional monitoring \citep{Pribulla2012,Sybilski2010}\footnote{http://www.projectsolaris.eu/} of this system that leads to precise timing measurements and additional information is suggested in order to unveil the true nature of the possible companions.

\acknowledgments{We would like to express our gratitude to the anonemous referee. Also we would like to thank Professor Chun-Hwey Kim and his students at Chungbuk National University and Dr. Chung-Uk Lee for fruitful discussions on LITE and dynamical aspects of circumbinary companions. TCH acknowledges KASI travel grant 2014-1-400-06. JH gratefully acknowledges the financial support of the University of Southern Queensland's Strategic Research Fund "STARWINDS Project". Part of the results presented in this work were carried out on the KMTNet computing cluster.}


\end{document}